\documentclass{aa}
\usepackage{graphics}
\input{psfig}
\begin{document}

\thesaurus{06     
           (03.11.1;  
            16.06.1;  
            19.06.1;  
            19.37.1;  
            19.53.1;  
            19.63.1)} 
\title{Dynamical Masses of Young Star Clusters in NGC
4038/4039\thanks{Based on observations collected at the European
Southern Observatory, Chile}}

\titlerunning{Masses of Clusters in NGC 4038/4039}

\author{Sabine Mengel\thanks{\emph{Present Address:} Leiden Observatory, P.O.
Box 9513, NL-2300 RA Leiden, The Netherlands}, Matthew D. Lehnert,
Niranjan Thatte, and Reinhard Genzel} \authorrunning{Mengel et al.}

\offprints{S. Mengel}

\institute{Max-Planck-Institut f\"ur extraterrestrische Physik
           Giessenbachstra\ss e, D-85748 Garching, Germany\\
           email: mengel@strw.leidenuniv.nl, mlehnert, thatte, genzel@mpe.mpg.de}

\date{Received ............... ; accepted ............... }

\maketitle

\begin{abstract}

In order to estimate the masses of the compact, young star clusters in
the merging galaxy pair, NGC 4038/4039 (``the Antennae''), we have
obtained medium and high resolution spectroscopy using ISAAC on VLT-UT1
and UVES on VLT-UT2 of five such clusters.  The velocity dispersions
were estimated using the stellar absorption features of CO at 2.29
$\mu$m and metal absorption lines at around 8500 \AA, including lines of
the Calcium Triplet.  The size scales and light profiles were measured
from HST images.  From these data and assuming Virial equilibrium, we
estimated the masses of five clusters.  The resulting masses range from
6.5 $\times$ 10$^5$ to 4.7 $\times$ 10$^6$ M$_{\sun}$.  These masses are
large, factor of a few to more than 10 larger than the typical mass of a
globular cluster in the Milky Way.   The mass-to-light ratios for these clusters in
the V- and K-bands in comparison with stellar synthesis models suggest
that to first order the IMF slopes are approximately consistent with
Salpeter for a mass range of 0.1 to 100 M$_{\sun}$.  However, the
clusters show a significant range of possible IMF slopes or lower mass
cut-offs and that these variations may correlate with the interstellar
environment of the cluster.  Comparison with the results of
Fokker-Planck simulations of compact clusters with properties similar to
the clusters studied here, suggest that they are likely to be long-lived
and may lose a substantial fraction of their total mass. This mass loss
would make the star clusters obtain masses which are comparable
to the typical mass of a globular cluster.

\keywords{star clusters -- dynamical masses -- NGC 4038/4039 -- IMF}
\end{abstract}

\section{Introduction}

During the last years, many interacting and merging galaxies were
discovered to host large numbers of young star clusters which formed
during the merging process (\cite{Hetal92, Wetal93, Meurer95, Wetal97,
W99, Zepf99}).  The overall properties of these clusters suggest that
they could be the progenitors of the globular cluster populations seen
in normal nearby elliptical and spiral galaxies (e.g., \cite{ZA93}).
Such an hypothesis also has the attractive implication that if
ellipticals formed through the merger of two large spiral galaxies (as
suggested in the popular hierarchical merging model of galaxy formation
e.g., \cite{Kauffmann93}), these young clusters might evolve into the
red (supposedly metal rich) part of the globular cluster population of
elliptical galaxies when the merger is complete (e.g.
\cite{Schweizer01}).  However, this hypothesis needs to be tested by
determining the characteristics of both, individual young clusters and
the cluster population as a whole.

Globular clusters have a typical mass of 1-2$\times$10$^5$ M$_{\sun}$ and
a mass function which is log-normal (e.g., \cite{Harris91}).  The
population of clusters in NGC 4038/4039 (``the Antennae"), however, has a
power law luminosity function, and the same shape is also suggested for
the mass function (\cite{W99, ZhangFall99}).  Masses determined from
photometric data are as large as a few $\times$10$^6$ M$_{\sun}$ for
some of the clusters (\cite{ZhangFall99, Mengel00}) and the determined
ages span a large range (\cite{W99, Mengel00}).

At first glance, the population of globular clusters in the Milky Way
and those in the Antennae seem to have vastly different ensemble
characteristics.  For example, the most massive clusters in the Antennae
are at least a factor of few more massive than those in the Milky Way.
However, given the large number of clusters formed in the Antennae, it
seems reasonable that the mass function is sampled up to a high
upper mass, and moreover mass loss during the evolution of the young
clusters can be expected to play a role.  
Evolution over a Hubble time might convert the power law
cluster mass function into a log-normal one, if lower mass clusters were
dissolved preferentially during the evolution.  Dynamical models like
those, for example, of Chernoff \& Weinberg (1990) and Takahashi \&
Portegies Zwart (2000) provide theoretical support for the necessary
differential evolution from a power-law mass distribution function into
a log-normal distribution.

Dynamical cluster masses are less model dependent than those determined
from photometry only, or, at the very least, have a different set of
dependencies.  The dynamical mass is derived from the stellar velocity
dispersion in combination with the cluster size and light profile.
While the photometric mass estimates depend on the model assumed for the
star formation parameters (time-scale, age, IMF slope, lower and upper
limiting masses, metallicity, etc), the dynamical mass estimates rely
only on the validity of the Virial equilibrium (i.e., the potential is
due to the collective gravitational effects of individual stars --
self-gravitating -- and is changing only slowly with time), and the
assumed constancy of the M/L ratio within the cluster.  The comparison
of photometric and dynamical cluster masses (cluster M/L) constrains the
slope of the mass function and may reveal the presence or absence of low
mass stars.  The fraction of low mass stars influences the survival
probability of the cluster during a few Gyrs of evolution, as clusters
rich in low-mass stars are less prone to destruction (e.g., Takahashi \&
Portegies Zwart 2000).

More specifically, we have used high spectral resolution spectroscopy
conducted at the ESO VLT to estimate the stellar radial velocity
dispersion $\sigma$, and high spatial resolution imaging from archival
HST images (\cite{W99}) to estimate the size scales (e.g., the projected
half-light radius r$_{hp}$).  This results in M$_{dyn}$ for individual
clusters which can be estimated using the equation:

\vspace{-0.2cm}
\begin{equation}\label{mvirequation}
M=\frac{\eta \sigma^2 r_{hp}}{G}
\end{equation}
\vspace{-0.2cm}

Where $\eta$ is a constant that depends on the distribution of the
stellar density with radius, the mass-to-light ratio as a function of
radius, etc.  The only assumption that goes into Virial relations of
this form is that the cluster is self-gravitating and is roughly in
equilibrium over many crossing times (i.e., it is not rapidly collapsing
or expanding).  The calculations of Spitzer (1987) indicate that for a
wide range of models, $\eta$ $\approx$ 10.  Under the assumption of
isotropic orbits, $\eta$ differs from 3 (the pure Virial assumption)
since the ``gravitational radius" must be scaled to the projected
half-light radius (or some measurable scale) and because the measurement
of the velocity dispersion is not the central velocity dispersion but is
a measurement which is weighted by the light profile.  So differing mass
profiles have different values of $\eta$ appropriate for estimating
their dynamical masses.  For example, globular clusters in the Milky Way
exhibit a range of concentration parameters (logarithm of the ratio of
the tidal radius to the core radius) in a King model (King 1966) of
approximately 0.5 to 2.5 (e.g., Harris 1991).  Over this range of
concentrations, $\eta$ varies from about 9.7 to 5.6.  Hence, accurately
measuring the light profiles (and of course the assumption that the
light profile traces the mass profile) is crucial in estimating the
masses.

Recent attempts at estimating the dynamical masses of similar clusters
in other galaxies has met with some success.  Notably, \cite{HF96a} and
\cite{HF96} were able to estimate the masses of two luminous young star
clusters in the blue compact dwarf starburst galaxy NGC 1705 (1705-1)
and NGC 1569 (1569A).  They measured velocity dispersions for 1705-1 and
1569A of 11.4 km s$^{-1}$ and 15.7 km s$^{-1}$ respectively.  Under the
assumption that $\eta$ = 10, Sternberg (1998) estimated that these
clusters have masses of about 2.7 $\times$ 10$^5$ M$_{\sun}$ for 1705-1
and 1.1 $\times$ 10$^6$ M$_{\sun}$ for 1569A.  More recently, Smith \&
Gallagher (2001) have estimated the dynamical mass of the luminous
star-cluster in M82, M82-F.  Assuming $\eta$=10, they find a mass of
1.2$\pm$0.1$\times$ 10$^6$ M$_{\sun}$.  This mass estimate implies that
the mass-to-light ratio of M82-F is very extreme and requires either a
very flat mass function slope or a lack of low mass stars for a
Salpeter-like mass function slope.  Interestingly, this small sample of
clusters seems to require a range of IMFs to explain their mass-to-light
ratios.

The purpose of the present paper is to derive masses for a small sample
of young compact clusters in the Antennae galaxies - the nearest merger
- and to use these mass estimates to constrain the functional form of
the IMF in these clusters.  Since it is important that the velocity
dispersion be determined from the stellar component of the cluster, we
have undertaken a program to observe extinguished clusters in the K-band,
to take advantage of the strong CO absorption band-heads beyond 2.29
$\mu$m, and unextinguished clusters in the optical, to take advantage of
the strong absorption of the Calcium Triplet (CaT) and other metal lines
around 8500\AA.  These features are strong in atmospheres of red
supergiants which would be expected in large numbers in clusters with
young ages ($\approx$6-50 Myrs) and thus particularly well-suited for
estimating the masses of clusters like those in the Antennae.  Given
that previous investigators (e.g., \cite{HF96a, HF96}) have measured
velocity dispersions around 15 km s$^{-1}$ suggests that to conduct such
measurements would require a minimum resolution of R$\approx$9000.  Such
resolutions are now available using ISAAC on VLT-UT1 which with its
narrowest slit delivers R$\approx$9000 and UVES on VLT-UT2 which
delivers a resolution of $\approx$38,000 with a 1\arcsec\ wide slit.

\section{Observations}\label{observations}

\subsection{Moderate Spectral Resolution K-band Spectroscopy}\label{IRobssec}

\begin{figure}
\psfig{figure=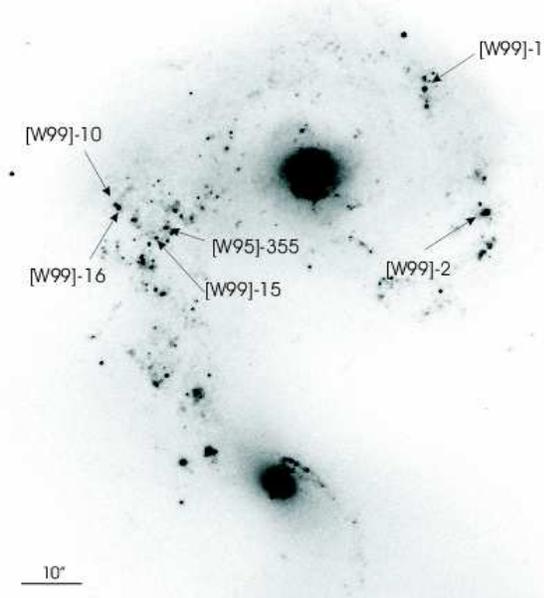,width=8.8cm,bbllx=40pt,bblly=240pt,bburx=490pt,bbury=710pt,clip=}
\caption[Clusterlocation]{ISAAC Ks-band image  of NGC 4038/4039 (taken during a run in April 2001) 
with labeling of the clusters that
were observed using ISAAC and UVES spectroscopy}
\label{clusterlocation}
\end{figure}

Spectroscopy in the K-band was performed with ISAAC at VLT-UT1 on April
17 2000.  The spectrograph was configured to have a 0\farcs3 slit and
the medium resolution grating was tilted such that the central
wavelength was 2.31$\mu$m with a total wavelength coverage from 2.25 to
2.37$\mu$m.  This was sufficient to include the $^{12}$CO (2-0) and
(3-1) absorption bands and at a resolution ($\lambda/\Delta\lambda$) of
about 9000.  Observations of late-type supergiant stars were taken so
that they could be used as templates for the determination of the
velocity dispersions\footnote{This research has made use of the SIMBAD
database, operated at CDS, Strasbourg, France}.  Various templates were
observed, however, due to incorrect classifications in the literature,
only one of them actually was an M-type supergiant.  More template stars
were observed by Linda Tacconi during an observing run on August 15 2000
using the same instrumental set-up.  Observations were performed by
nodding along the slit and dithering the source position from one
exposure to the next.  Atmospheric calibrator stars were observed
several times during the night.

The target clusters were selected using the results from NTT-SOFI
imaging observations performed in May 1999 (Mengel, Ph.  D.  Thesis, der
Ludwig-Maximilians-Universit\"at M\"unchen, 2001; Mengel et al.  2001, in
preparation).  They included narrow band imaging in the CO(2-0)
band-head and revealed those clusters which are at an age where the
near-infrared emission is dominated by red supergiants.  The clusters
with the highest equivalent width in the CO band-heads, highest
luminosity, and high extinction were selected as the ISAAC targets
(for the location of all observed clusters see Fig.~\ref{clusterlocation}).  
Two target clusters were positioned in the first slit position.  These were
two clusters with a separation of $\sim 5$\arcsec\ located in the region
where the galaxy discs apparently overlap (``the overlap region").  One
of them has an optical counterpart, [W99]15, the other one is detectable
only in the I-band, therefore denoted as a ``Very Red Object (VRO)'' by
\cite{WS95} (with the designation:  [WS95]355).  Their detectability in
at least one band from the HST imaging program of \cite{W99} was a
necessary additional constraint, since the compactness of each cluster
means that their radii must be measured from HST data (see \S
\ref{sizeestsection}).  The second slit position included only one
bright cluster ([W99]2), which has a relatively low extinction and was
also observed with UVES.  Due to the extreme narrowness of the slit,
object drifts made several re-acquisitions necessary, especially when
the object was transiting.  The ISAAC observations are summarized in
Table 2.

\begin{table*}
\begin{center}
\caption[Summary of ISAAC observations]{Summary of ISAAC spectroscopy.
The first slit position included two clusters, the second one cluster.
The supergiants were observed to provide template spectra for the
determination of velocity dispersions.  The slit width was 0\farcs3,
resulting in a resolution R = $\lambda/\Delta\lambda$ of 9000. 
Seeing variations are expected to have no effect on the spectral
resolution given the narrow slit width. 
The total integration time was split up into several integrations of
duration T$_{int}$.  The K-band magnitudes were determined from an ISAAC
Ks-band image (aperture diameter 1\farcs74).  The seeing value refers to
an average for the monitored V-band seeing during the integration.  The
stars were observed during two observing runs (2000 and 2001), and those
marked with LT were kindly observed by Linda Tacconi during another run,
because the 2000 run had resulted in only one useful supergiant
spectrum.}

\begin{tabular}{llccccccl}\hline\hline
Position & Object     & m$_V$  & m$_K$ & m$_{K_0}$ & T$_{int}$ & Total ON time & Seeing & Comments\\
	&	     & [mag]  & [mag]	&   [mag]   &[s]	& [s] & & \\\hline
Slit 1  &[W99]15     & 19.4	& 15.9 &15.8 & 600 & 15600 & 0\farcs7 & Difficulties keeping both \\
        &[WS95]355   & $\approx$22.1	& 15.7 & 15.4 & 600 & 15600 & & targets in slit during 2 integrations\\
Slit 2  &[W99]2      & 17.0	& 14.0 & 14.0 &600 & 2400s & 0\farcs9 &\\\hline
\end{tabular}
\begin{tabular}{l}
and the following stars: K5Ib (HD183589), K5Ib (HD53177, LT), M0I (CD-44 11324, LT), \\
M0-1Iab (HD 42475), M1I (BD+0 4030), M1Iab (HD 98817), M1Ib (HD 163755), M2Iab (SAO 15241), \\
M3Ia-Iab (HD103052), M3.5I (BD+13 1212), M5Iab (HD 142154)\\\hline\hline
\end{tabular}
\end{center}
\label{ISAACobs}
\end{table*}

\subsection{High-Resolution Optical Spectroscopy}\label{optspectsection}

High-resolution optical echelle spectroscopy was performed using UVES at
VLT-UT2 on the night of 18 April 2000.  The instrument was configured
with a slit width of 1\arcsec, which resulted in a resolution of
R$\approx$38,000 (depends on wavelength).  A dichroic was placed in the
light path allowing the use of both the red and blue arms of the
spectrograph.  However, in this paper we will discuss only the results
obtained with data from the red arm.  The central wavelength of the red
arm was shifted to 8400\AA\ since part of the CaT (at NGC 4038/4039
redshift of z$\sim$0.005 located roughly at:  8540\AA, 8585\AA\ and
8705\AA) would have fallen right in the small gap between the two CCDs
that cover the lower and the upper part of the spectra in the red arm.
The CCDs were read out with a binning of 2x2, 50kHz and high gain.  For
the cluster selection we applied the same criteria as for the clusters
observed with ISAAC, however, for the UVES observations, we selected
only those with relatively low optical extinction and bright I-band
magnitudes.  To boost the efficiency of the observing, we additionally
required the clusters to have a nearby cluster within the slit of UVES
(which is only 10\arcsec \ due to the echelle design).  The clusters
observed were [W99]1 and [W99]43 in one slit position, [W99]10 and
[W99]16 in a second slit position.  Unfortunately, [W99]43 and [W99]10
turned out to be too faint to extract a spectrum of sufficient S/N for
further analysis.  A third slit position covered [W99]2, common to ISAAC
and UVES.  Obtaining two independent observations of [W99]2 provides an
independent estimate of the velocity dispersion to test the
uncertainties of our measurements and to gauge whether there might be
some systematic differences in velocity dispersions determined from the
CaT in the optical with UVES versus that obtained from the CO band-head
with ISAAC in the near-IR.  A range of template supergiants was observed
(early K through late M-type supergiant stars), but also several hot
main sequence stars (late O through mid B-type main sequence stars).
Massive main sequence stars are expected to dominate the blue spectrum
and to contribute significantly to the flux in the I-band (Bruzual \&
Charlot 1993).  Due to the high dispersion and faint sky background,
integration times could be very long and were only limited by the desire
to keep the number of cosmic ray hits down to a reasonable limit.  The
UVES observations are summarized in Table~2.

\begin{table*}
\begin{center}
\caption[Summary of UVES observations]{Summary of UVES spectroscopy of
science targets (star clusters and template stars).  The spectral
resolution was R$\approx$38,000, and differences in the seeing for different
objects had no measurable effect on the shape of the absorption features in
the stellar spectra.}

\begin{tabular}{llccccccl}\hline\hline
Position & Object & m$_V$ & m$_K$ & m$_{K_0}$ & T$_{int}$ & T$_{tot}$ & Seeing & Comments\\
        &       & [mag] &  [mag] & [mag] & [s] &   [s]     &       &       \\\hline
Slit 1 & [W99]1 & 17.6 & 14.7	& 14.7	& 4800s & 9600s & 0\farcs7 & required pausing of integration\\
       & [WS95]43 & 20.0 &	&	& 4800s & 9600s &  & due to A.O. software crash\\
Slit 2 & [W99]10 & 19.0 & 15.6	& 15.6	& 4800s+1800s& 11400s & 0\farcs85 & required pausing of integration \\
        & [W99]16 & 19.0 & 15.5 & 15.5  & 4800s+1800s & 11400s & & due to A.O. software crash\\
Slit 3 & [W99]2 & 17.0 & 14.0	& 14.0	& 4600s & 4600s & 1\farcs2 &\\\hline
\end{tabular}

\begin{tabular}{l}
And the following stars:  M3Iab (HD 303250), M5I (HD 142154), M1Ib (HR
6693), MIa-Ib (HR 4064)\\
K7IIa (HD 181475), K5Ib (HR 7412), K5II (HR 7873), K3Ib-IICN (HR 6959),
K3II (HR 6842)\\
K2.5IIb (HR 7604), K2II (HR 6498), K2Ib (BM Sco), K0.5IIb (HR 6392),
K0Ia (HR 6392)\\
B5V (HR 5914), B5V (HR 5839), B2V (HR 6028), B2V (HR 6946) with integration
times\\
between 1s and 20s, and some atmospheric and flux calibrators\\\hline\hline
\end{tabular}
\end{center}
\label{UVESobserv}
\end{table*}

\section{Reduction and  Analysis}

\subsection{Reduction of the near-IR data}\label{IRredandanalsection}

\begin{figure*}
\psfig{figure=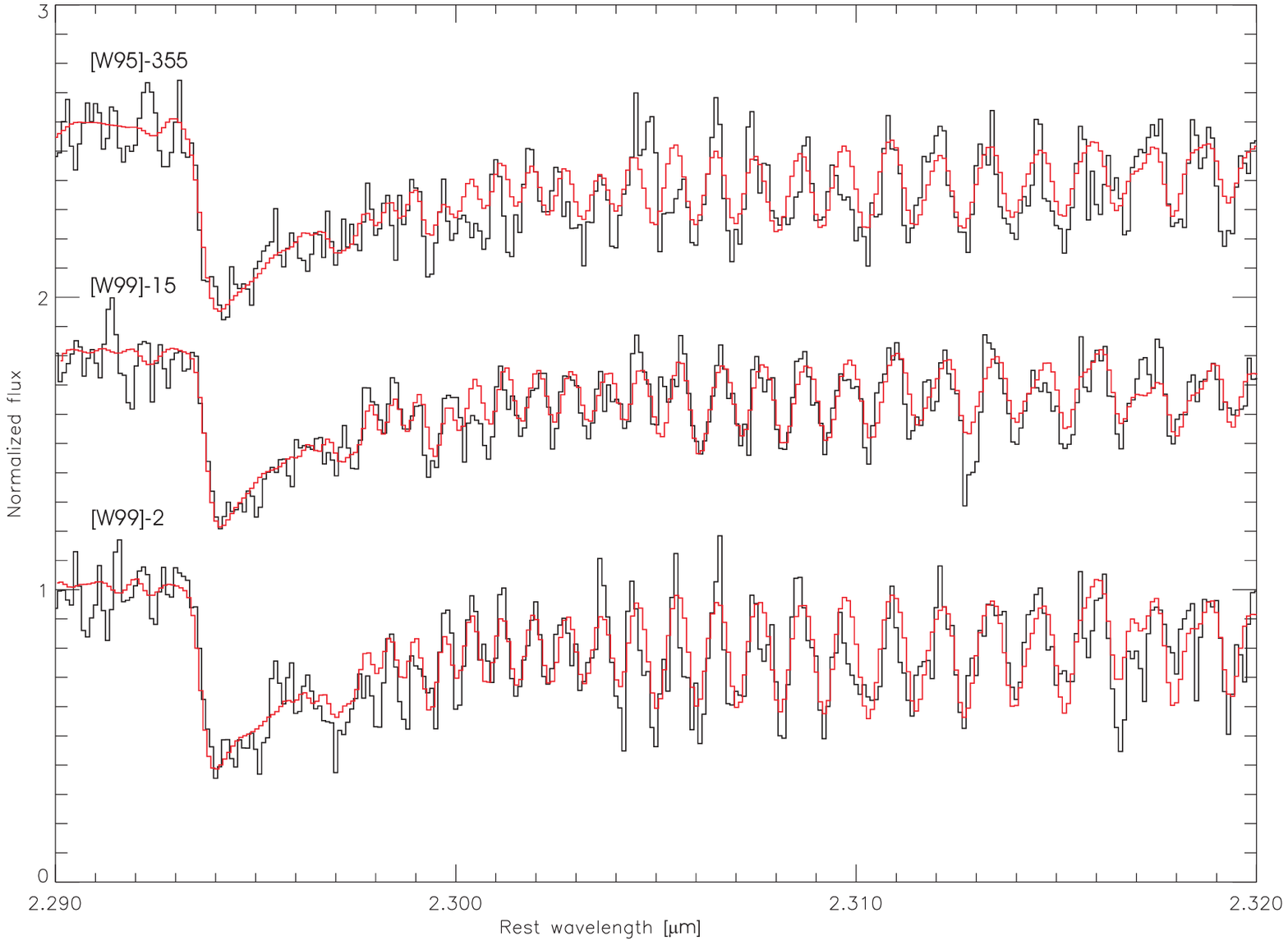,width=17.6cm,height=10cm,bbllx=32pt,bblly=400pt,bburx=540pt,bbury=796pt,clip=}
\psfig{figure=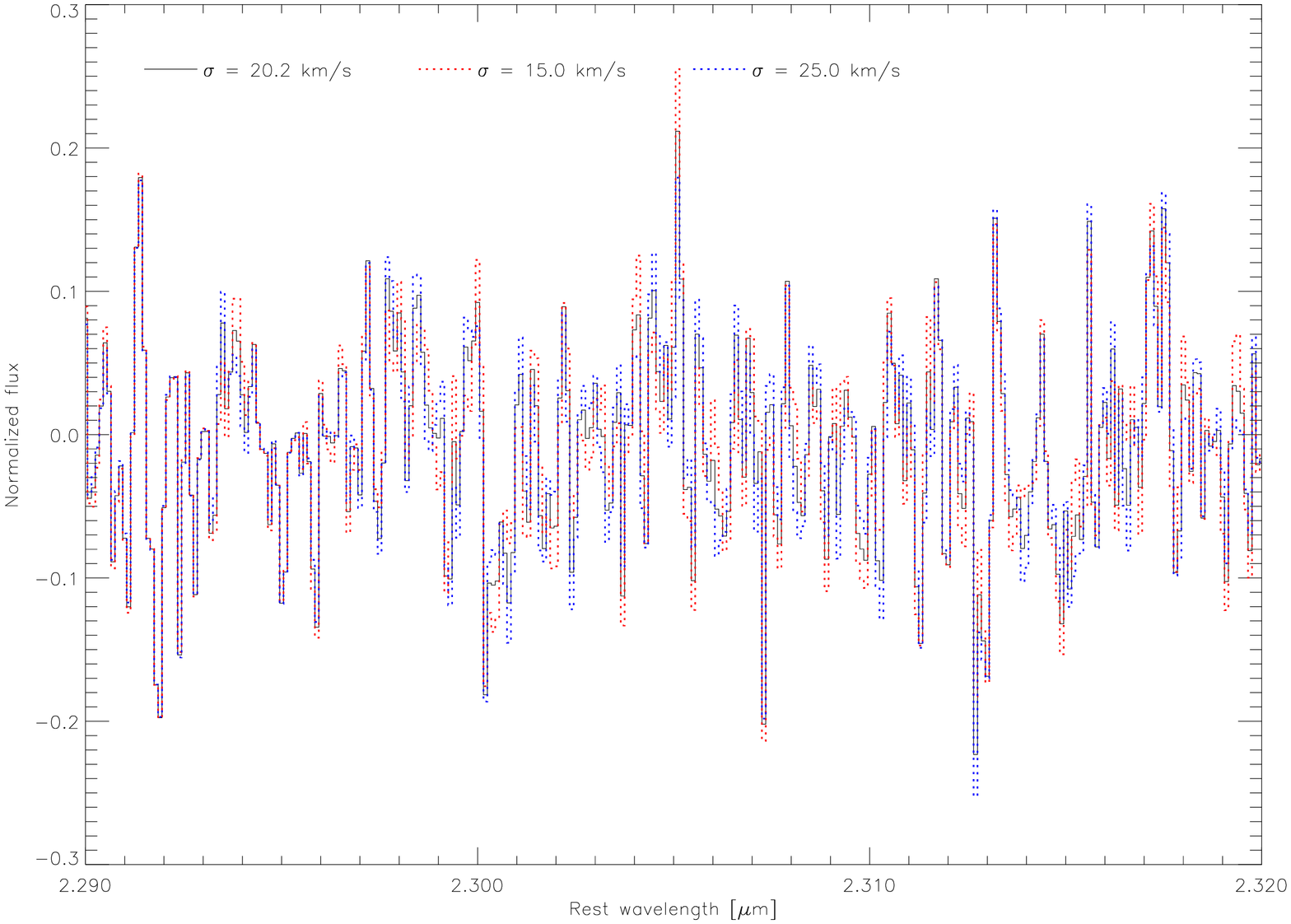,width=17.6cm,height=7.5cm,angle=90}
\caption[ISAAC spectra of three clusters with fits]{ISAAC Spectra of the
three clusters [WS95]355, [W99]15 and [W99]2 (top
to bottom, black) with best fits (grey). The bottom plot shows the residuals of the fit
to spectrum [W99]15 for the best fit and for two fits with lower and higher  velocity dispersions.
This gives a visual impression of how much the velocity dispersion affects the spectral features.}
\label{bestfitISAAC}
\end{figure*}

The reduction of ISAAC data was performed in a standard way using the
IRAF data reduction package\footnote{IRAF is distributed by the National
Optical Astronomy Observatories, which are operated by the Association
of Universities for Research in Astronomy, Inc., under cooperative
agreement with the National Science Foundation.}.  The reduction
included dark subtraction and flat-fielding (using a normalized
flat-field created from dome flat observations) on the two-dimensional
array.  Sky subtraction was performed pairwise, followed by a rejection
of cosmic ray hits and bad pixels.  The spectra were then corrected for
tilt and slit curvature by tracing the peak of the stellar spatial
profile along the dispersion direction and fitting a polynomial to the
function of displacement versus wavelength.  The corresponding procedure
was also applied to the orthogonal direction.  Single integrations were
combined by shifting-and-adding, including a rejection of highest and
lowest pixels.  The rejection could not be applied to observations of
[WS99]2, because they consisted of only a limited number of single
frames.  The object spectra were then extracted from user defined
apertures.  A linear fit to the background (below $\approx$7\% of the 
peak intensity for all clusters) on both sides of the object
spectrum was subtracted.

The spectra were wavelength calibrated using observations of arc discharge
lamps.  An atmospheric calibrator (B5V) was observed and reduced in the
same way as the target and used to divide out the atmospheric absorption
features from the spectra.

\subsection{Reduction of the optical echelle data}\label{optredandanalsection}

Reduction of the UVES data was performed within the echelle spectra
reduction environment of IRAF.  Reduction included bias subtraction,
flat-fielding and extracting the spectrum from user supplied apertures.
A background is fit to neighboring regions and subtracted.  
Also here, the background was below $\approx$7\% of the 
peak intensity for all clusters. This step
includes a bad pixel rejection.  The full moon was fairly close to the
Antennae during the integrations, and its scattered light introduced a
solar spectrum into the data.  However, this contribution is very
effectively removed during background subtraction.

Wavelength calibration required the identification of many lines in the
ThAr-spectra for each of the echelle orders which were identified using
a line table available from the ESO-UVES web pages.  A dispersion
function was then fit and applied to the object data.  This data set
contains each order of the spectrum in a separate channel.  These
separate channels are combined into one final spectrum covering the
total wavelength range.  This involves re-gridding of the wavelength
axis into equidistant bins and averaging in the overlapping edges of the
orders.  The spectral resolution obtained is R$\approx$38,000 across all
orders.

\subsection{Estimating the Velocity Dispersions}\label{estveldispsection}

For each spectrum (both the optical and near-IR spectra), we estimated
the $\sigma$ of the broadening function (assumed to be a Gaussian) which
best fit the cluster spectrum in the following way.  The stellar
spectrum (the template spectrum) was broadened with Gaussian functions
of variable $\sigma$ in velocity space.  The resulting set of spectra
were then compared with the cluster spectrum.  The best fit was
determined by evaluating $\chi^2$ and then search for the minimum of the
function $\chi^2$(v$_r$,$\sigma$) using a simplex downhill algorithm for
the tour through parameter space.

\subsubsection{Unique Character of the near-IR Estimates}
\label{nearIRestsection}

Obviously, the template spectrum has to be a good overall match to the
cluster spectrum, otherwise an erroneous velocity width will be derived
for the cluster.  For a star cluster that formed $\sim$10 Myrs ago, late
K through early M supergiants are expected to provide the largest
contribution to the flux at 2.3$\mu$m.  However, according to population
synthesis models (e.g., \cite{Letal99}, \cite{Sternberg98}), there will
be a non-negligible contribution to the flux from hot main sequence
stars.  Since the stars are hot (O and B-type stars), this ``diluting
continuum" will be an essentially featureless continuum which solely
decreases the equivalent width of the CO band-heads.  This has the
effect of shifting the apparent dominating stellar type towards higher
effective temperatures.  Starting out with a template spectrum with weak
CO features leads to very low velocity dispersions, the opposite is the
case if an M5I star (strong band-heads) is used, with vast differences
in the results (a few km s$^{-1}$ to up to about 30 km s$^{-1}$).

To first order, the velocity dispersion determined from different
stellar templates agrees if the CO equivalent widths match.  This is
achieved by adding a (positive or negative) continuum to the stellar
spectrum, which dilutes or enhances the CO features.

To second order, there are two different indicators that can be used to
constrain the matching stellar template.  Both relate to the shape of
the CO band-head, due to variations in that shape between the different
supergiant spectral classes.  The various rotational transitions are
resolved in the $^{12}$CO(2-0) and the $^{12}$CO(3-1) band-head.  We
performed tests of our analysis technique by creating a simulated
cluster spectrum with an input velocity dispersion $\sigma_{in}$, added
noise (usually S/N = 15) and re-determined the velocity dispersion
$\sigma_{out}$ using different template star spectra.  With the correct
template (out = in), we obtained in 30 test runs an average
$\sigma_{out}$ = 14.7$\pm$0.7 km s$^{-1}$ (with $\sigma_{in}$ = 15.0 km
s$^{-1}$).  Performing the same test on an input spectrum that was
diluted by a flat continuum (10\%), but still fitting with the undiluted
template, the velocity dispersion increased (17.9$\pm1.7$ km s$^{-1}$),
as expected.  Fitting other stellar types (M0I, M1I) yields similar
results.  But the mismatch can be diagnosed by:

\begin{itemize}
\item{determining $\sigma$ over several different wavelength ranges.  With
the matching template, $\sigma_{out}$ depends only weakly (maximum
differences $\sim 6\%$) on the selected wavelength range, no matter if
only the band-head, only the overtones, or both are covered.  A template
mismatch causes differences in the $\sigma_{out}$ for different
wavelength ranges of $\sim 25\%$ typically, but sometimes more.}
\item{inspecting the shape of the fit around the tip and the rising edge
of the band-head, which provides a good match only for the correct template}
\end{itemize}

\subsubsection{Unique Character of the Optical Estimates}\label{optestsection}

The determination of the velocity dispersion for the optical echelle
data used the same procedure as for the ISAAC spectra, relying on the
Calcium Triplet around 8500\AA, but also using the Mg absorption feature
at 8800\AA\ and other weaker metal absorption lines between 8400 and
9000\AA.

Examining the spectra in detail, it was obvious that they are composed
of a mixture of light from supergiants and hot main sequence stars.
This means that the spectra cannot be simply fit by broadening the
stellar template, but that either the hot star contribution needs to be
added to the template, as was done in the case of the ISAAC spectra, or
that the contribution needs to be subtracted from the cluster spectrum.

Here, the procedure was different from that applied to ISAAC data,
because the hot main sequence stars were not completely featureless over
the analyzed wavelength range, but had some Paschen absorption features.
Pa16, Pa15 and Pa13 lie slightly red-ward of the CaT wavelengths 8498\AA,
8542\AA, and 8662\AA, respectively.  Since the widths of these lines are
mainly due to line broadening in the atmosphere and rotation of the star
itself and only an insignificant fraction of the width is due to the
velocity dispersion of the star cluster, they cannot be used to estimate
the velocity dispersion.  Therefore this weak contribution was
subtracted from each cluster before the velocity dispersion was
estimated.  The amount of subtraction was estimated by the strengths of
the Paschen line absorption and the appropriate template for the
subtraction was chosen for its ability to remove the contribution
accurately.  Given the S/N of the spectra and the weakness of the
features removed, the final characteristics of the subtracted spectrum
were not very sensitive to the exact template used.

\begin{figure}
\begin{minipage}{8.8cm}
\psfig{figure=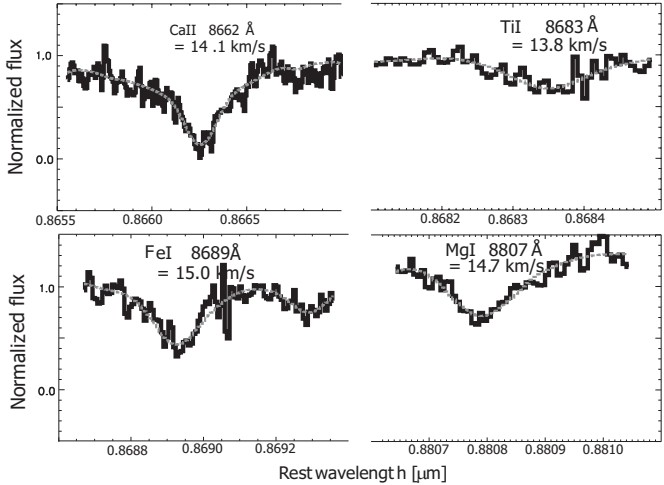,width=8.8cm}
\caption[Fit to the UVES spectrum of the cluster W99-2]{Displayed are
parts of the normalized spectrum of cluster [W99]2 (black), together
with the fit (red).  The four pieces were fit separately, and the
average value of the velocity dispersion is $\sigma$ = 14.2 km s$^{-1}$.
It is worth noting that this is the cluster which was also observed with
ISAAC, and to compare the results (see Figure~\ref{bestfitISAAC}).}
\label{cl9uvesfit}
\end{minipage}
\end{figure}

The cluster [W99]2 merits some specific discussion since it has both
ISAAC and UVES observations and also some peculiarities were noted
compared to the other clusters observed with UVES.
Fig.~\ref{cl9uvesfit} shows fits of an M3I spectrum to pieces of the
spectrum of the cluster [W99]2.  The latter had a 45\% contribution of
B2V stars subtracted.  This contribution was determined by interactively
subtracting the spectrum of the B2V star until the Paschen absorption
features were reduced down to the noise level of the cluster spectrum.
In addition, we fit the continuum of the cluster spectrum using a
combination of a B2V and an M3I star spectrum, in which the relative
contributions of the two template stars was the fit parameter.  The best
fitting combination again included a 45\% contribution of B2V stars.
Moreover, the relative contributions of M3I and B2V stars in I and K
band agree with the determined contributions to the ISAAC and UVES
spectra, respectively, for this cluster.

\subsection{Estimating the Sizes of the Observed Clusters}\label{sizeestsection}

In order to make an estimate of the mass, it is necessary to have a
measure of the cluster radius and profile shape.  Whitmore et al.
(1999) found that many of the young star clusters in images taken with
the Wide Field Planetary Camera on HST (WFPC-2) of the Antennae were
slightly-to-well-resolved and estimated their sizes.  Their main aim was
to determine the average radii of the clusters and to compare them to
those observed in other galaxies, both to the young clusters in other
mergers and to globular clusters.  They found that the mean effective
radius of the clusters in the Antennae is r$_{eff}$ $\approx$ 4~pc.
However, Whitmore et al.  (1999) did not present their results on
individual clusters.  Brad Whitmore kindly provided us with their
estimates of the effective radii for the clusters we observed with ISAAC
and UVES.  But given the importance of size and light profile to the
mass determination, we additionally measured them ourselves from the HST
I-band images.

Similar to what Whitmore et al.  (1999) did to make such estimates we
reduced archival HST I-band images (see Whitmore et al.  1999 for the
details concerning the images) using the drizzle technique (see the
Drizzle Handbook available from STScI).  Given the pixel sizes of the
HST chips (0\farcs101 pix$^{-1}$ for the Wide Field Camera arrays and
0\farcs045 pix$^{-1}$ for the Planetary Camera array), and the distance
to NGC 4038/4039, which implies a scale of 93 pc arcsec$^{-1}$, the
factor of two improvement in the sampling of the HST
point-spread-function (PSF) due to drizzling the images is critical in
getting a robust estimate of the sizes.  At the typical cluster size,
r$_{eff} \approx$ 4~pc, the subtended scale is only 0\farcs043 or about
one pixel of the Planetary Camera!  For this analysis we used the {\it
ishape} routine described by \cite{Larsen99} implemented in the BAOLAB
data reduction package developed by Larsen.  The {\it ishape} routines
were especially designed to work more efficiently than typical
deconvolution routines at determining the sizes of slightly extended
objects.  {\it ishape} convolves the user-provided PSF with the given
cluster profile and determines the minimum $\chi^2$ for a range of sizes
using a simplex downhill algorithm to find the minimum.  The PSF we
selected for this convolution is one from the Tiny Tim program (Krist
1995) which Whitmore et al.  (1999) argue persuasively is appropriate
for these data.  We fit each cluster with a range of provided profiles
which include Gaussian profiles, King profiles of several concentration
parameters (\cite{King66}), and Moffat profiles of two different
concentrations (Moffat 1967).  The program produces a reduced $\chi^2$
to measure the significance of the fit and various images (artificial
cluster image, the original input image, an image of the residuals
between original and fit, and an image of the weights applied to each
pixel) to judge the quality of the fit.  The code also provides a handy
weighting procedure that enables the rejection of ``outliers" in the
radial cluster profile, thereby excluding, for example, neighboring
stars and/or clusters.

\begin{table*}
\begin{center}
\caption[Cluster masses]{The cluster masses as they were derived from
the ISAAC and UVES spectra (in column ``Inst.'' indicated by I and U,
respectively) in comparison with the supergiant template spectrum.  The
age was derived from the combination of W$_{Br\gamma}$, W$_{CO}$, and
W$_{CaT}$. For cluster [W99]2, it agrees well with the age estimated
from the UV spectrum of 7$\pm$1 Myrs \cite{W99}.  The size is the projected half-light radius r$_{hp}$
estimated as described in \S \ref{sizeestsection} (for the cluster marked with
$^a$, we could not obtain a satisfactory fit and used the value provided by
B. Whitmore). W$_{CO}$ was estimated from the
ISAAC spectra between rest-frame wavelengths 2.2924$\mu$m and
2.2977$\mu$m, ([W99]2 also from 3D integral field spectroscopy), and
W$_{CaT}$ from the UVES spectra, according to the definition in
\cite{Diaz89}.  The stellar velocity dispersion $\sigma$ was determined
as described in \S \ref{estveldispsection}, for [W99]2 the average
value of ISAAC and UVES measurements is given ($\sigma$=14.0$\pm 0.8$
and 14.3 $\pm 0.5$ km s$^{-1}$, respectively).  M$_{vir}$ is the Virial
mass determined from equation \ref{mvirequation} with the 1$\sigma$-uncertainties
given in column $\sigma_M$ .  The light-to-mass
ratios were derived from the extinction corrected magnitudes and a
distance modulus to NGC 4038/4039 of 31.41.}

\begin{tabular}{lcccccccccc}
\hline
\hline
Cluster & Inst. & W$_{CO}$ & W$_{CaT}$ & Age & $\sigma$ & r$_{hp}$ & log M$_{vir}$ & $\sigma_M$ & log L$_V$/M &log L$_K$/M \\
        & I/U   & [\AA] & [\AA] & [10$^6$yr] & [km/s] & [pc] & [M$_{\sun}$] & [\%] & [L$_{\sun}$/M$_{\sun}$] & [L$_{\sun}$/M$_{\sun}$]  \\\hline
$[$WS95$]$355   & I & 16.3$\pm$0.2 & - & 8.5$\pm$0.3 & 21.4$\pm$0.7 & 4.8$\pm$0.5$^a$ & 6.67 & 12 & ... & 1.05 \\
$[$W99$]$15  & I & 17.0$\pm$0.2 & - & 8.7$\pm$0.3 & 20.2$\pm$0.7 & 3.6$\pm$0.5 & 6.52 & 16 & 0.89 & 1.09 \\
$[$W99$]$2   & I/U & 16.2$\pm$0.2 & 6.4$\pm$1.0 & 6.6$\pm$0.3 & 14.2$\pm$0.4 & 4.5$\pm$0.5 & 6.31 & 12 & 1.54 & 2.00 \\
$[$W99$]$1   & U & 17.5$\pm$1  & 8.6$\pm$1.0 & 8.1 $\pm$0.5 & 9.1$\pm$0.6 & 3.6$\pm$0.5 & 5.81 & 19 & 1.76 & 2.22 \\
$[$W99$]$16  & U & 19$\pm$4  & 9.9$\pm$1.0 & 10$\pm$2  & 15.8$\pm$1 & 6.0$\pm$0.5 & 6.51 & 15 & 0.50 & 1.20 \\\hline

\end{tabular}
\end{center}
\label{massestable}
\end{table*}

\section{Results}\label{resultssection}

Our results are summarized in Table~3.  We find a range
of cluster velocity dispersions from about 9 km s$^{-1}$ to over 20 km
s$^{-1}$.  One concern about any such measurements is the reliability of
the final results.  For the optical echelle data, the instrumental
resolution, $\sigma_{instrument}$=3.2 km s$^{-1}$ ensures that the
cluster line profiles are always well-resolved.  The situation for the
near-IR data is less clear.  For the clusters where the
velocity dispersions were estimated from the near-IR CO band-head, the
measured values are above or approximately at the instrumental
resolution ($\sigma_{instrument}$=14.2 km s$^{-1}$) and are thus likely
to be secure.  However, we were worried about possible systematics
effects that might influence the final estimates.  To make an estimate
of these possible problems, we constructed ``test data'' from the
stellar templates that were artificially broadened and had noise added
to mimic the real cluster data from both the ISAAC and UVES
observations.  This allowed us to investigate possible systematic
problems in the way each of the data sets was analysed and allowed us to
discover the most robust ways of analyzing the data.  In addition, for
one of the clusters, [W99]2, we obtained both UVES and ISAAC spectra of
this cluster to observe any systematic differences between the two data
sets that might lead us to conclude that there are systematic
differences in the properties among the clusters.

\begin{figure}
\begin{minipage}{8.8cm}
\begin{minipage}{8.8cm}
\psfig{figure=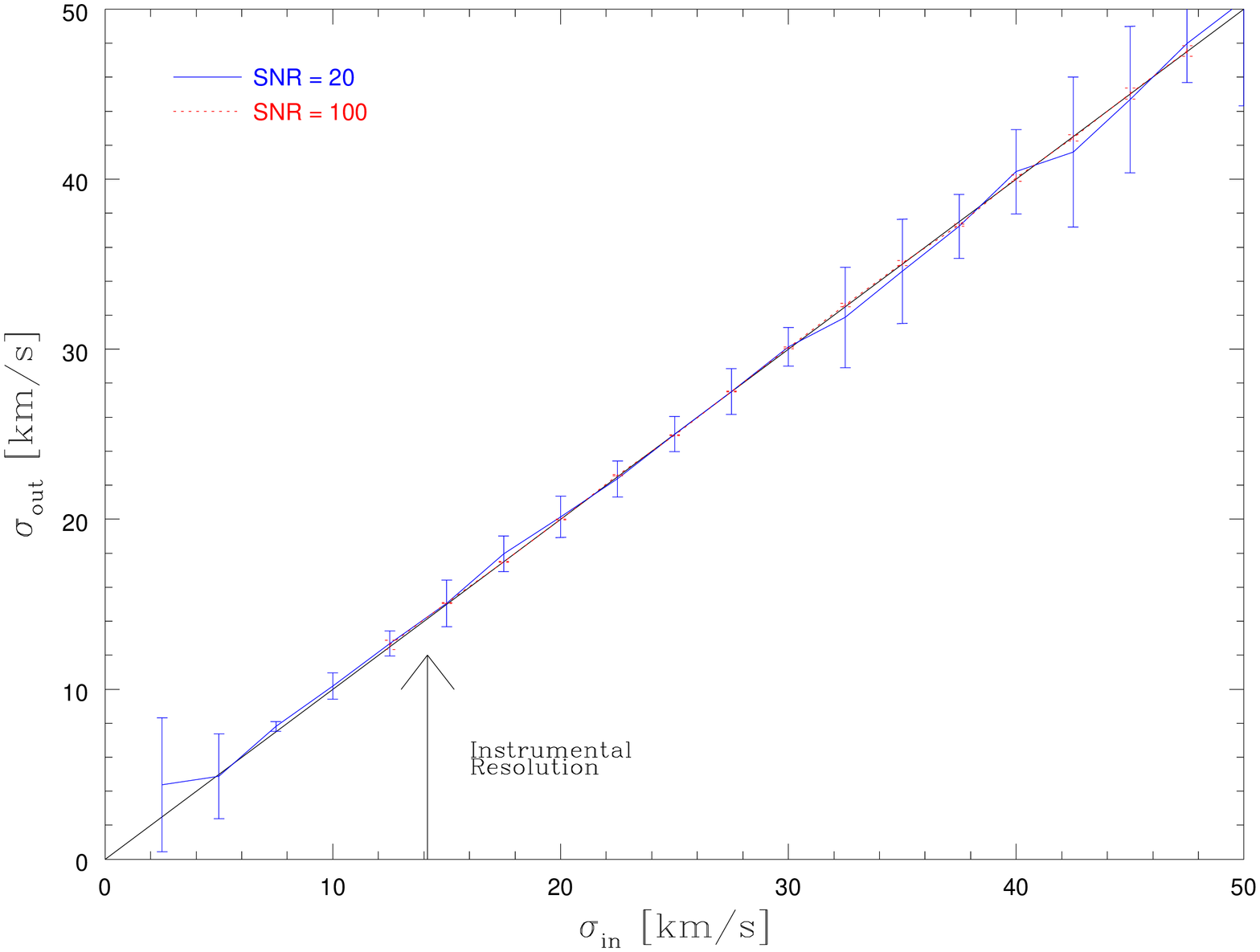,width=8.8cm,angle=-90}
\end{minipage}
\vfill
\begin{minipage}{8.8cm}
\psfig{figure=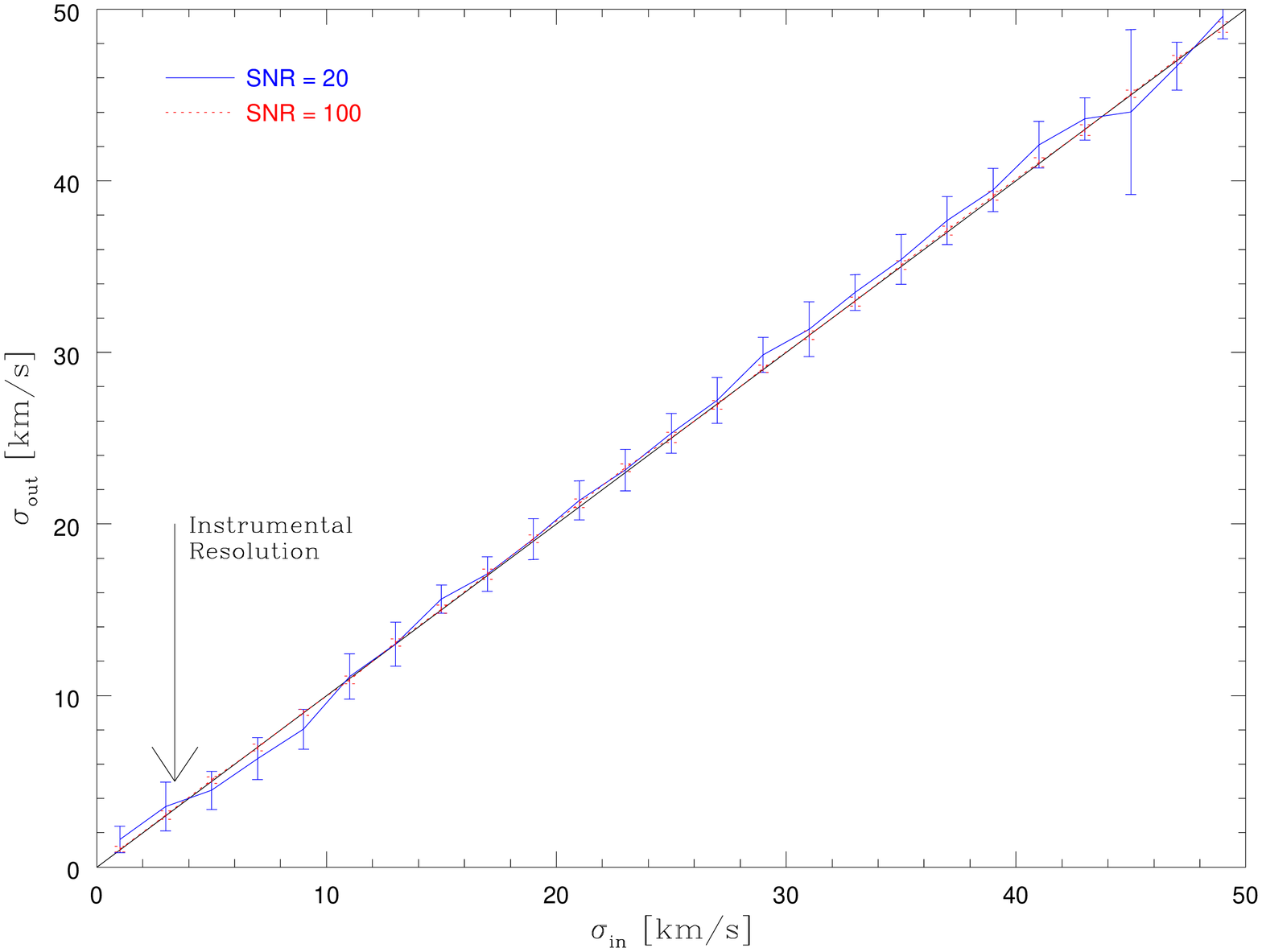,width=8.8cm,angle=-90}
\end{minipage}
\caption{Artificial data with a S/N comparable to our data (S/N = 20)
and with a S/N = 100
were created from stellar template spectra and the input velocity dispersion
was redetermined. The average value of 30 test runs with the variance is
displayed for both instruments (Top: ISAAC, Bottom: UVES).}
\label{sigmainout}
\end{minipage}
\end{figure}

From experiments with the test data, we found no systematic problems in
the estimated dispersions as long as the template was a good match (as
described earlier) and for velocity dispersions that were above
approximately 50\% of the instrumental resolution.  We conducted these
simulations for the range of S/N spanned by the data for the clusters
(Figure~\ref{sigmainout}).  As a by product of these simulations, we
were able to estimate the uncertainties in our estimates by conducting
many (several tens) trials for each input broadening width.  From the
distribution of resulting dispersion estimates, we were able to
determine the average output velocity dispersion to look for systematic
offsets and the variance to provide a likelihood estimate of the
uncertainty in any one measurement.

\begin{figure*}
\begin{minipage}{8.8cm}
\psfig{figure=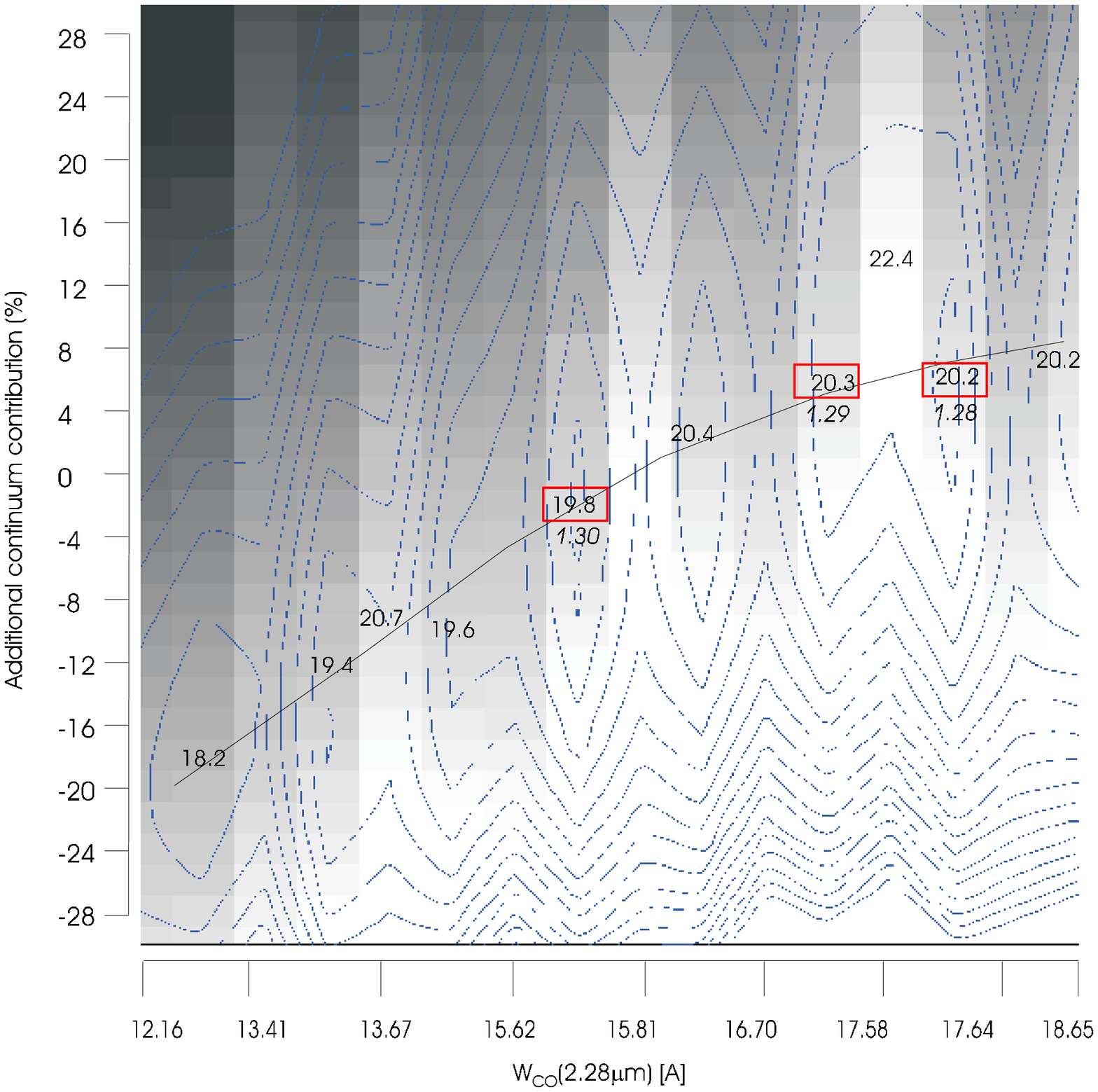,width=8.8cm}
\end{minipage}
\hfill
\begin{minipage}{8.8cm}
\psfig{figure=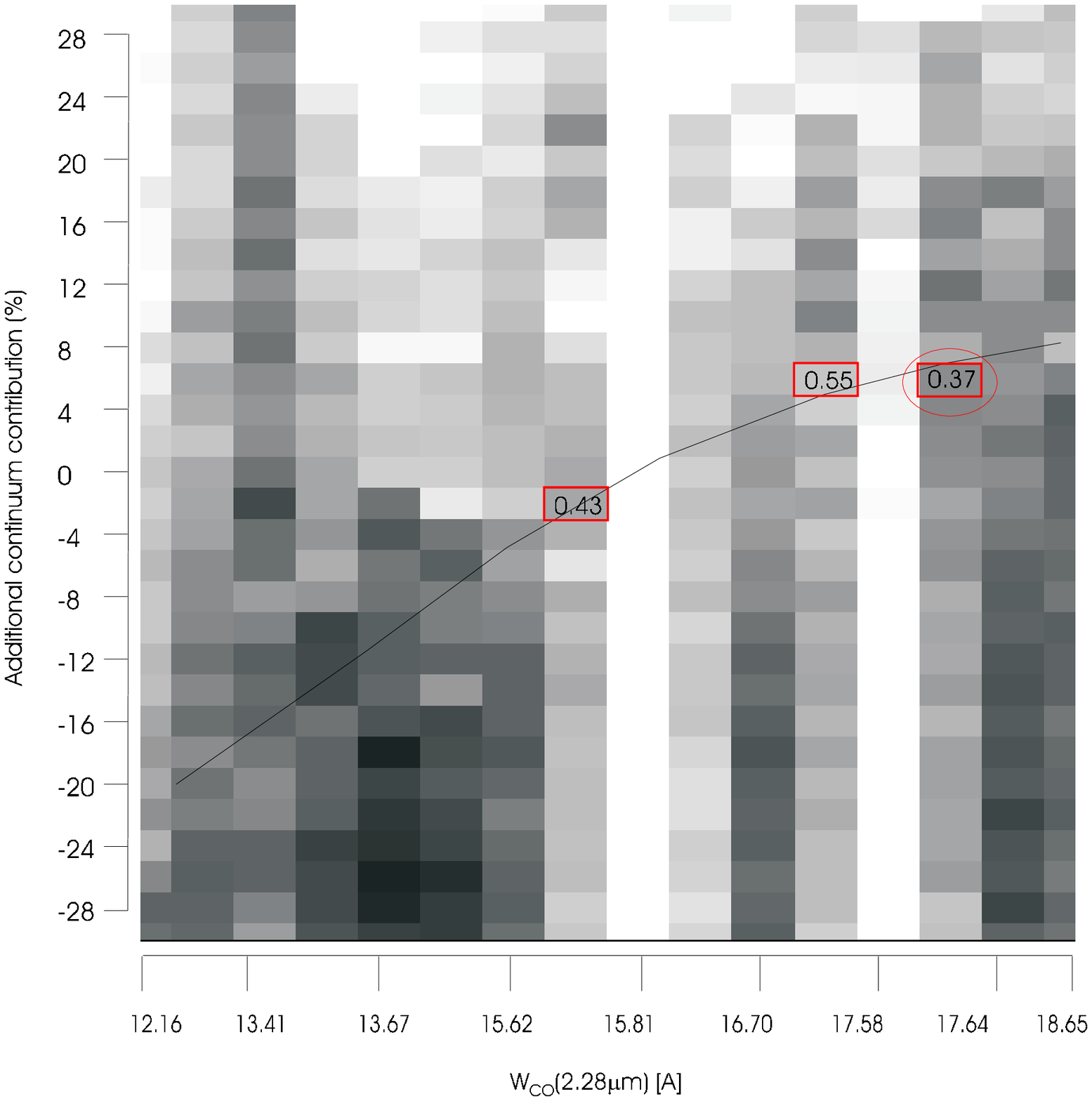,width=8.8cm}
\end{minipage}
\caption[]{Visualization of the velocity dispersion determination for cluster
W99-15. \\
{\bf Left:}  Greyscale shows the velocity dispersion (black $\approx$10 km\,s$^{-1}$
to white $\approx$27 km\,s$^{-1}$), contours show the corresponding $\chi^2$ (from
1.28 to 2.0).  Arranged along the x-axis are different input supergiant
template stars, arranged according
to increasing CO equivalent width. The input spectra between labelled
columns were created by averaging two neighbouring spectra. The y-axis
arranges the template spectra by added or subtracted continuum
contribution, ranging from -30\% to +30\%. The line was added to guide the
eye along the regions of lowest $\chi^2$, which depends mainly on the CO
equivalent width of the template spectrum matching that of the
cluster. Some of the $\sigma$-values for the lowest $\chi^2$ in a column
are labelled, and the three best fitting $\sigma$ values are boxed, with
the corresponding $\chi^2$ value listed below the box. 
{\bf Right:} The best fitting $\sigma$ value is chosen from the three
$\sigma$ values with the lowest $\chi^2$ (they are marked with boxes in
the left figure) by requiring that $\sigma$ depend only weakly on the
selected wavelength range. The corresponding diagnostic map is shown,
which displays the standard deviation in the velocity dispersion (grey
scale, ranging between 0.05 (black) and 1.63 km\,s$^{-1}$ (white)). Out of
the pre-selected three best fits, the third spectrum, located between the
columns labelled ``17.58'' and ``17.64'' shows the lowest variations in
velocity dispersion for the five wavelength regions. The numbers in the
boxes give the standard deviation for the velocity dispersions for the
three best cases. Thus, the best matching template (marked by an
ellipse) has a W$_{\rm CO}$ of 17.6 \AA\ and a continuum contribution of
6\%, resulting in a velocity dispersion of 20.2 km\,s$^{-1}$, with a
standard deviation of 0.37 km\,s$^{-1}$.
}
\label{sigmadetw15}
\end{figure*}

We applied the three criteria described in \S~\ref{nearIRestsection} to
our cluster data by creating a stellar template grid with CO(2.28$\mu$m)
equivalent widths between 12.2\AA\ and 18.7\AA\ (we used the nine
template stars listed in Table~2 and created intermediate
spectra by averaging of two spectra), and added continuum levels of
-30\% to +30\%.  One of these maps is displayed in
Figure~\ref{sigmadetw15}.  The greyscale image of velocity dispersions
has a contour overlay of the corresponding $\chi^2$, and the minimum
positions are to first order located at regions of comparable CO
equivalent widths.  We selected for the average $\sigma$ value only
those template spectra which had a weak dependence of $\sigma$ on
selected wavelength range, which is displayed on the second map in
Figure~\ref{sigmadetw15}, and additionally verified by visual inspection
that the shape of the fit around the band-head is well-matched.

The best-fitting template for cluster [W99]15, for example, is a star with
a CO equivalent width of 17.6\AA, with an additional 6\% continuum
contribution.  This matches very closely the predictions of stellar
contributions from evolutionary synthesis models:  For example,
according to STARS (\cite{Sternberg98}), the K-band light in a cluster
of 8.5 Myrs is composed of 95\% M1/M2 type supergiants, $\approx3$\% hot
main sequence stars and $\approx$1.5\% A-G type supergiants.

The ISAAC and UVES observations gave an estimate of $\sigma$=14.0$\pm$0.8 
and 14.3$\pm$0.5 km s$^{-1}$ for [W99]2, respectively, where the
uncertainties in each measurement are the combination of the
uncertainties in a single measurement as determined from the test data,
and the deviations between different templates and fit wavelengths
ranges.  The difference in the estimated velocity dispersions is only
0.3 km s$^{-1}$.  This difference is well within the random error we
have estimated using the test data and demonstrates that there are
unlikely to be any systematic differences in using the UVES and ISAAC
data (or the I-band compared to K-band or the CaT plus weaker lines
compared to the CO band-heads) to estimate the velocity dispersions.

We found that the best fitting models for the light profiles of the
clusters were King models with tidal to core radii ratios of 15
(c=1.176) or 30 (c=1.477; both concentrations typically provided fits
that were of indistinguishably good statistical significance but clearly
better than the other fitted models, i.e., Gaussian or Moffat profiles).
The effective radii of the observed clusters were approximately 4 pc,
similar to what Whitmore et al.  (1999) estimated as the average radius
of the population of clusters in the Antennae.  So while the selected
clusters are among the most luminous in the Antennae, they are certainly
not among the largest.

\subsection{Estimating the Metallicities and Ages of the Clusters}

Given the wide wavelength coverage of the UVES data, some interesting
features are contained in the spectra.  One such feature is the MgI line
at 8806.8\AA.  This feature has been shown by \cite{Diaz89} to be
sensitive to the metallicity.  The strength of this feature
in two clusters is between 0.6 and 0.8\AA,  which according to the
empirical calibration of the equivalent width of the MgI feature
suggests approximately solar metallicity (there is significant scatter
in the relationship between metallicity and line strength).  However,
the equivalent width of the MgI absorption feature in [W99]2 is
$\approx$1.2\AA, which indicates super-solar metallicity.  Unfortunately,
for the clusters with only ISAAC near-infrared spectroscopy, it is not
possible to estimate the metallicities of the clusters from these data
alone.

To determine the ages of the clusters, we followed the prescription
outlined in Mengel et al. (2001), with the additional information
from the CaT.  This method relies on the equivalent widths of the
CO-band head at 2.29$\mu$m, CaT lines, and the Br$\gamma$ emission line
(where available for each cluster).  The Br$\gamma$ emission strength
estimates come from Mengel et al. (2001) or Mengel et al. (2001, in
preparation) with the later estimates resulting from narrow band Br$\gamma$
imaging of the Antennae.  We used the diagnostic plots from Leitherer et
al.  (1999) of the equivalent widths of these features versus age
assuming an instantaneous burst, solar metallicity (but twice solar for
cluster [W99]2), and a Salpeter IMF between 1 and 100 M$_{\sun}$.  The
strength of using these features is that the equivalent
width versus age relations have different dependencies as a function of
age.  Br$\gamma$ is strong in the early evolution of the cluster and
weakens on the main sequence evolutionary time scale of the most massive
stars with significant ionizing photon output (ages less than about 6-8
million years).  The CaT lines reach significant equivalent widths at
about 6 million years (as the I-band light becomes dominated by red
supergiant stars) while the CO band-head at 2.29$\mu$m becomes
significant at about 8 million years (as the K-band light becomes
dominated by red supergiants).  The combination of these diagnostics
have a useful and relatively unambiguous age sensitivity to
instantaneous burst populations over the age range from 0 to 20 million
years, beyond which, the age determination ceases to be unambiguous.

From these line strengths we estimated the ages given in
Table~3, ranging from about 6 to 10 million years, which reflects the
selection criterion of strong CO absorption, required for the
determination of the velocity dispersion (as discussed in
\S~\ref{observations}).  Moreover, we note that the metallicity
estimates can play a crucial role in determining the ages.  As we
discussed for [W99]2, we have evidence that it has a super-solar
metallicity.  In Mengel et al.  (2001), we found that [W99]2 exhibits
strong CO band-heads as well as relatively strong Br$\gamma$ emission
and discussed the age determination of this cluster under the assumption
of solar metallicity.  Mengel et al.  (2001) found that one age was
difficult to reconcile with the measurements and speculated that the
simultaneous Br$\gamma$ and strong CO band head could be reconciled with
a single age if the low spatial resolution of our data allow for
different physical regions of different ages to contribute to the final
spectrum, the cluster had an extended duration of star-formation, or
that shocks were present.  Our finding that this cluster may have
super-solar metallicity releaves this dilemma in that it is now possible
to reconcile all of the measurements in that strong CO absorption now
appears at an earlier age (at an age where Br$\gamma$ emission is also
present).  In this case, we now find the age of 6.6 Myrs instead of 8.1
Myrs under the assumption of solar metallicity.  Further we note, in our
current sample, [W99]2 is that only cluster that shows significant
Br$\gamma$ emission and CO 2.29$\mu$m lines simultaneously.

\section{Discussion and Implications}\label{discussion}

\subsection{Are the Clusters Bound and Have They Undergone Mass Segregation?}\label{bounddiscussion}

The underlying assumption of any mass estimate using the velocity
dispersion will be that the cluster is self-gravitating (i.e., bound).
From the velocity dispersions and half-light radii we estimate that the
crossing times ($t_{cross} \sim r_{hp}/\sigma$) range from about 1 to a
few $\times$10$^5$ yrs.  We determined the ages of these clusters to be
around 8 Myrs and thus these clusters have already survived for 20-50
crossing times.  It is clear that at least initially, these clusters
must have been gravitationally bound, and that the structure of these
clusters is likely to be changing slowly over many crossing times.
This assumption is also supported by N-body simulations performed by
Portegies Zwart et al. (1999) on the evolution of a cluster with a similar
concentration like the ones in our study. Core radius and density stay
almost constant between 4 and 10 Myrs, the endpoint of their study.
Only one IMF was used (Scalo), but we assume that these findings should be
valid unless the IMF slope is very shallow (below $\alpha \approx 1.5$).

The stars are expected to be segregated by mass on the ``two-body
relaxation time-scale''.  Given that we have an estimate of the velocity
dispersion and mass densities, it is possible to estimate the two-body
relaxation time-scale.  Using the arguments in Spitzer \& Hart (1971) we
estimate that the half-light relaxation time scale is several 100 Myrs.
Since the cluster ages are much less than this, dynamical mass
segregation seems unlikely.  However, several very young star clusters,
like R136 in 30 Doradus (age around 4 Myrs, see e.g.  Brandl et al.,
1996) show evidence for mass segregation, which suggests that either the
more massive stars formed preferentially near the center, or that the
segregation occurs on a faster time-scale:  N-body calculations by
Portegies Zwart (2000) achieved some mass segregation within a few
crossing times.  If mass segregation was present in the Antennae star
clusters, both the half mass radius and the velocity dispersion, and
therefore the mass, would be underestimated by our calculations.

\subsection{The Estimates of the Masses}\label{massdiscussion}

Since we have argued that these clusters must have been gravitationally
bound initially (although subsequent dynamical evolution and mass loss
through massive star winds and supernova explosions may unbind the
clusters), we can now use the Virial theorem, the light profiles (and
under the assumption that the light distribution follows the mass
distribution, i.e., the mass-to-light ratio is not a function of
radius), and the velocity dispersion to estimate the masses of the
clusters.  The masses were derived assuming 1) that the best fitting
profile was a King model (the light profiles were well fit with King
models with concentrations of 1.176 or 1.477) 2) that the mass
distribution follows the light profile, and 3) that the measured
velocity dispersions were in fact not the central velocity dispersions
but were projected velocity dispersion weighted by the projected light
profile.  Again, the corrections between the half-light radius and the
core radius (defined in the King model) and central velocity dispersion
and the measured velocity dispersion were given by those from the King
models themselves (see, for example, McLaughlin 2000 for these
relationships).  The masses we derive are listed in
Table~3 and range from about 6.5 $\times$10$^5$
M$_{\sun}$ to 4.7$\times$10$^6$ M$_{\sun}$.  The uncertainties in the
mass estimates given in Table~\ref{massestable} only represent the
statistical uncertainties estimated for the velocity dispersion and
half-light radius and do not represent any additional uncertainties due
systematic errors (as might be caused by a radial dependence on the
mass-to-light ratio, systematically over-estimating the size of the
cluster, erroneous functional form of the light profile, etc).

\begin{figure}
\begin{minipage}{8.8cm}
\psfig{figure=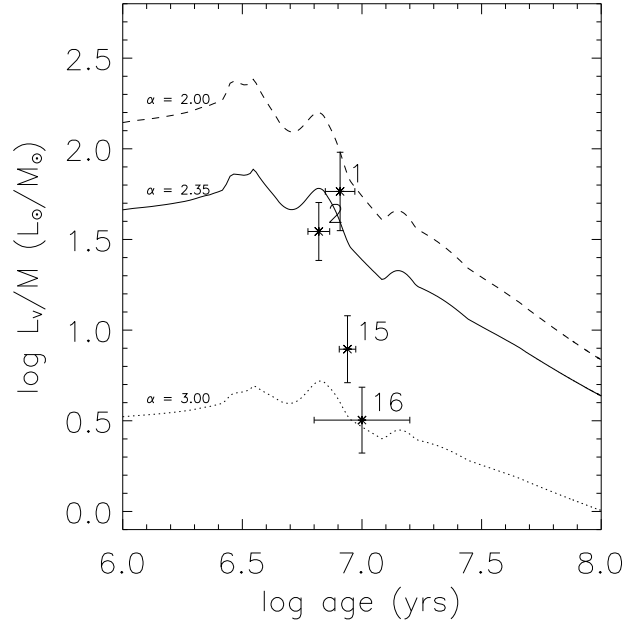,width=8.8cm}
\hspace{0cm}
\caption[Mass-to-light for the V-band]{The V-band light-to-mass ratios
for the young compact clusters in the Antennae compared with models of
Leitherer et al.  (1999) for an instantaneous burst of total mass 10$^6$
M$_{\sun}$ with a power-law slope, solar metallicity , and a stellar
mass range from 0.1 to 100 M$_{\sun}$.  The power-law slopes are 2.35
(solid line), 2.00 (dashed line), and 3.00 (dotted line).  The points
represent the data for the antennae clusters with V-band photometry from
HST and with the extinction estimated as stated in the text.
Unfortunately, cluster [WS95]355 is too faint to be detected in
the HST WFPC-2 images.  The error bars represent a combination of the
error in the mass and the uncertainty in the photometry and extinction
estimate was assumed to be 10\%.  The light-to-mass ratios shown take
into account the mass lost through stellar winds and supernova over
time.}
\label{LoverMV}
\end{minipage}
\end{figure}

We caution that while we believe the fits to the light profiles are
robust, given that the HST data used to estimate the size scales has a
resolution that is close to the measured half-light radii, it is
possible that we have systematically over-estimated the cluster sizes.
This could lead to a systematic over-estimate of the total dynamical
masses by the same proportion.  The relation between cluster
concentration and mass is non-linear.  For our typically best fitting
models (c=1.176 or c=1.477), both lower or higher values would lead to a
decrease in mass (by 2\% if c=0.477 instead of 1.176, and by 25\% if c=2
instead of 1.477).  This means that unless the clusters are much more
concentrated than estimated, the masses can be overestimated only
slightly.

\subsection{Mass-to-Light Ratios of the Clusters and Implications for the IMF}\label{MLIMFdiscussion}

One of the critical parameters to understanding star-formation in and
the dynamical evolution of these compact, young star-clusters is their
initial mass function.  If we find that these clusters are best
described as having an IMF with significant numbers of low mass stars
(say Salpeter slope down to 0.1 M$_{\sun}$), then the dynamical
evolution will be driven through the ejection of these numerous low mass
stars by binaries and massive stars in the cluster.  Moreover, the
fraction of the total mass lost through stellar winds and supernova
explosion will be relatively small (10s of percent; Leitherer et al.
1999).  It is then likely, since the amount of overall ejected mass is
relatively small compared to the total mass, for the clusters to remain
bound and long-lived.  If however, there is a lack or relative deficit
of low mass stars (either through an IMF that is truncated at the low
mass limit or that has a relatively flat slope), then the total mass
loss from the cluster due to stellar ejection (since higher mass stars
would then be more likely to get ejected) and the fraction of the total
mass lost through stellar winds and supernova would also be
proportionally higher both of which would tend to increase the
likelihood that cluster becomes unbound (e.g., Chernoff \& Weinberg 1990;
Takahashi \& Portegies Zwart 2000).

\begin{figure}
\begin{minipage}{8.8cm}
\psfig{figure=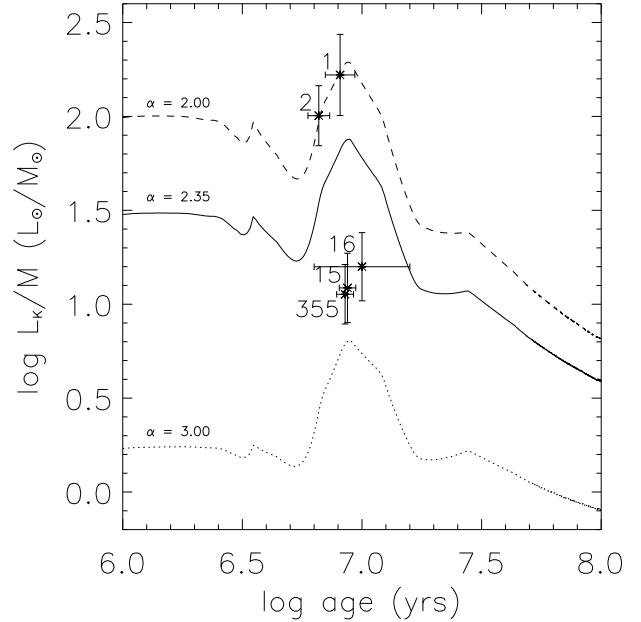,width=8.8cm}
\hspace{0cm}
\caption[Mass-to-light for the K-band]{The K-band light-to-mass ratios
for the young compact clusters in the Antennae compared with models of
Leitherer et al.  (1999) for an instantaneous burst of total mass 10$^6$
M$_{\sun}$ with a power-law slope, solar metallicity (we did not
add the track for 2$\times$Z$_{\sun}$, which would be appropriate for 
cluster [W99]2because the difference
is below 0.1 dex, well within our uncertainties), and a mass range
from 0.1 to 100 M$_{\sun}$.  The power-law slopes are 2.35 (solid line),
2.00 (dashed line), and 3.00 (dotted line).  The points represent the
data for the Antennae clusters with ground-based K-band photometry
(Mengel et al.  2001, in preparation) and with the extinction estimated
as stated in the text.  The error bars represent a combination of the
error in the mass and the uncertainty in the photometry and extinction
estimate was assumed to be 10\%.  The model light-to-mass ratios shown
take into account the mass lost through stellar winds and supernovae over
time.}
\end{minipage}
\label{LoverMK}
\end{figure}

In the following we use the derived masses and ages, in combination with
the photometry and extinction estimates, to constrain the IMFs of these
clusters through the use of population synthesis models (see for
example, Sternberg 1998).  The approach uses a comparison of the cluster
with population synthesis models to predict the mass-to-light ratios as
a function of the cluster age.  In our case, we can perform this
comparison in both the V-band and the K-band, which is useful because
these bands differ in their M/L-evolution with age and in their
extinction sensitivities.  For example, the K-band is less sensitive to
the extinction than the V-band (by about a factor of 10) and also shows
a strong dependence on age through the significant effect on the K-band
luminosity when supergiants become prevalent in the cluster stellar
population.  There are also practical differences in the two bands.  The
K-band data are ground-based and hence are sensitive to our ability to
remove the background from the photometry of the clusters and also might
include contributions to their total magnitudes from other nearby
clusters.  The V-band data are HST WFPC-2 images and hence are less
sensitive to contamination and background subtraction uncertainties.
Although this last difference is partially mitigated by the fact that
our K-band image used for this analysis has a seeing of $\approx$0\farcs4 to
0\farcs5.

Keeping these differences and limitations in mind, we attempt to
constrain the functional form of the IMF in Figures~6 and 7.  In the
plot of the log L$_K$/M versus age, the most striking feature is
that the clusters form two groups, several clusters have L$_K$/M at
their estimated ages that seem best described with an IMF having a
relatively flat slope, or perhaps slightly steeper slopes, but truncated
low mass limits (higher than the 0.1 M$_{\sun}$ in the model).  For
example, the difference in mass between a Salpeter IMF slope with an
initial mass range from 0.1 -- 100 M$_{\sun}$ compared to a Salpeter IMF
slope with an initial mass range from 1.0 -- 100 M$_{\sun}$ is a factor
of 2.6.  Such a difference would shift the Salpeter IMF slope line in
Figures~6 and 7 upwards by 0.4 dex implying that a Salpeter IMF slope
with an initial mass range from 1.0 -- 100 M$_{\sun}$ is consistent with
the results for clusters [W99]1 and [W99]2.  The results for clusters
[W99]15 and [WS95]355 are more consistent with a steeper IMF slope,
approximately 2.5 with an initial stellar mass range of 0.1 -- 100
M$_{\sun}$ or a steeper slope with a truncated lower mass.  In the
light-to-mass ratios in the V-band, this difference is less clear.  The
light-to-mass ratios for [W99]1 and [W99]15 give consistent results in
both the K- and V-bands.

Similar results for young, compact clusters in nearby galaxies have also
been presented.  Sternberg (1998) using data from the literature for two
clusters, one in NGC 1705 (NGC1705-1) and another in NGC 1569
(NGC1569-A), found that the light-to-mass ratio of these two clusters
indicated rather shallow IMF slopes ($\approx$ 2.5 for NGC1569-A and $<$
2 for NGC1705-1 if their IMFs extend down to 0.1 M$_{\sun}$).  Smith \&
Gallagher (2001) found in their analysis of M82-F that it either
requires an extremely flat IMF slope ($\alpha$ $<<$ 2.0) or that it has
a steeper IMF slope ($\alpha$ = 2.3) but then has a lower mass cut-off
of about 2-3 M$_{\sun}$.  However, some caution is necessary in the
interpretation of M82-F.  The photometry of this cluster is rather
uncertain given that it was heavily saturated in the HST images used by
Smith \& Gallagher (2001) to estimate its magnitude and the extinction
(E$_{B-V}$ =0.9$\pm$0.1).  Smith \& Gallagher (2001) also re-analysed
some of the half-light radius and magnitude measurements of NGC1705-1
and NGC1569-A and although the detailed estimates of the light-to-mass
ratios for NGC1705-1 differ, they come to the same conclusions as
Sternberg (1998).  We show a graphical comparison of our results with
those obtained for M82-F, NGC1705-1, and NGC1569-A from Smith \&
Gallagher (2001) in Fig.~8.

\begin{figure}
\begin{minipage}{8.8cm}
\psfig{figure=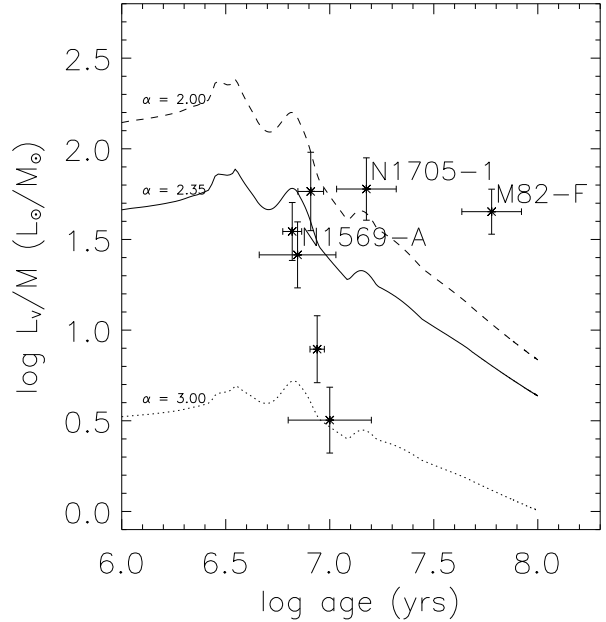,width=8.8cm}
\hspace{0cm}
\caption[Mass-to-light for additional clusters in the V-band]{The
points are the same as in Fig.~\ref{LoverMV} but now with additional
results from Smith \& Gallagher (2001) for the clusters M82-F,
NGC1705-1, and NGC1569-A.  The results obtained for the clusters
NGC1705-1 and NGC1569-A are similar to those in Sternberg (1998)
except for NGC1705-1 which in Smith \& Gallagher (2001) has
a L$_V$/M lower by a factor of approximately two mainly due to
their re-evaluation of the cluster half-light radius (revising the
size upward by about a factor of 2).}
\end{minipage}
\label{litLoverM}
\end{figure}

If the segregation of the cluster points in Figure~7 is a real effect,
it means that the IMF varies significantly between clusters.  We do not
yet have the possibility to disentangle lower mass cutoff and IMF slope,
but the two different groups of clusters differ in their ratio of low-
to high mass stars by approximately a factor 5 (``low'' and ``high''
relating to stars below and above 3 M$_\odot$, respectively).  

There has been substantial debate in the literature as to whether or not
different environments lead to different forms of the IMF.  While there
does not exist a thorough understanding of the process of
star-formation, possible mechanisms that have been investigated for
determining the characteristics of the IMF are variations in the local
thermal Jeans mass (e.g., Larson 1985; Klessen \& Burkert 2001),
turbulence (e.g., Elmegreen 1997), gravitationally-driven fragmentation
(e.g., Larson 1985), cloud-cloud and star-star coalescence (e.g., Nakano
1966; Bonnell, Bate, \& Zinnecker 1998) and several other processes (see
for example, \cite{Larson99, Elmegreenrev99} and references therein).
While it would be highly speculative to suggest that the range of IMF
slopes and/or low mass cut-offs we observe supports or refutes any of
these hypotheses for what physical processes are responsible for
determining the characteristics of the IMF, it is useful to discuss what
constraints these results might offer in understanding the origin of the
IMF.

There are two attractive hypotheses for explaining the observed segregation.  
One is that perhaps the difference in the relative number of low mass stars is
related to the environment in which the cluster is born.  Clusters
[W99]15 and [WS95]355 suffer greater extinction on average than most
clusters in the Antennae and both lie in the over-lap region of the two
merging disks.  Clusters [W99]1 and [W99]2 lie in the ``western loop''
of the Antennae's northern most galaxy, NGC4038 and have relatively low
extinctions (A$_V$$\approx$0.3 magnitudes).  In spite of the clusters
having similar ages, the environments of [W99]15 and [WS95]355 appear to
be much more rich in dust and gas (Whitmore \& Schweizer 1995; Wilson et
al.  2000; Mengel et al.  2001).  However, cluster [W99]16, while
physically lying near the clusters [W99]15 and [WS95]355, appears to
suffer much less extinction.  Presumably, since the region where
[W99]15, [W99]16, and [WS95]355 were formed is also the region where the
two disks are interacting, the gas pressure (both the thermal and
turbulent) would be high due to dynamical interactions of the two
interstellar media. The local thermal Jeans mass, which is considered to
play a role in setting the mass scale for low mass stars (e.g., Larson 1999), 
is proportional to T$^2$P$^{-1/2}$, where T and P are the
total pressure and temperature (e.g., Spitzer 1978). Clouds
with low temperature and/or high pressures would form clusters with IMFs
weighted towards low mass stars, which would be applicable for 
the more extinguished clusters.  Conversely, clouds
that are collapsing in relatively low extinction, gas poor, more
quiescent regions might be influenced by the ionizing radiation of
nearby young clusters raising its gas temperature.  Such a situation
would favor the formation of clusters with IMFs that have relatively
large fractions of high mass stars. Such an idea
might also explain the mass functions of NGC1569-A and NGC1705-1 whose
relatively low metallicities could favor high gas temperatures (by reduced
cooling and high background radiation fields).
Thus studies of young compact clusters in the Antennae
and other galaxies appear to agree with the expectation from the low
mass scaling set by the Jeans mass.

The second plausible hypothesis is that in Fig.~7 and 8 the
clusters with relatively large ratios of light-to-mass are also among
the lower mass clusters.  N1569-A, N1705-1, M82-F, [W99]1, and [W99]2
all have masses less than about 2 $\times$ 10$^6$ M$_{\sun}$.  Clusters
[W99]15, [W99]16, [WS95]355 all have masses greater than 3 $\times$
10$^6$ M$_{\sun}$.
While the total mass of a cluster influencing the
observed IMF has been discussed in the literature, it is most commonly
discussed within the context of random sampling of stars in clusters
with relatively few stars (e.g., Elmegreen 1999).  The Antennae clusters
are certainly not in this limit since they have many $\times$ 10$^5$
stars.  However, it may be that there is a process internal to the
cluster that inhibits preferentially the formation of low mass stars and
that this may scale roughly with the mass of the cluster.  One
possibility is that the massive stars through radiation pressure might
inhibit or shut off the formation of low mass stars.  

Obviously using compact, young star clusters in the Antennae and other
galaxies to provide observational constraints on the processes that lead
to the characteristics of the IMF will require much more observational
and theoretical work.  However, even the relatively sparse and modest
results provided here and in the literature are tantalizing and show the
promise of this type of observation in advancing our understanding of
the process of cluster and star-formation.

\subsection{The Evolution of the Most Massive Clusters:  Will They
Dissolve?}

Are these young clusters progenitors of the old globular cluster systems
that we observe in normal galaxies?  If they are, how would the
population of young clusters need to evolve to have a mass distribution
that resembles that of the old globular cluster system?  These
fundamental questions can only be addressed when we have characterized
the masses, initial mass functions, and strength and change in the
gravitational potential in which these clusters form and evolve.

Various models have been constructed to follow the evolution of clusters
with structure similar to globular clusters (see \cite{spitzer87} and
references therein).  Perhaps the most relevant for this study are those
that employ the Fokker-Planck approximation in solving the dynamics of
the constituent cluster stars (\cite{cohn79,CW90, TPZ00}).  The dynamics
of Fokker-Planck models are relatively simple to calculate compared to
full N-body models and thus other astrophysical effects that might be
important to the dynamical evolution of the cluster can be included
(such as including a range of stellar masses distributed with a
realistic IMF, stellar mass loss, stellar evolution, etc).  While these
models have their limitations, the results of these studies provides a
useful guide on how these young, compact clusters of the Antennae might
evolve given our determined constraints on the shape of their IMFs and
concentrations.  The Fokker-Planck models (e.g., \cite{cohn79,CW90,
TPZ00}) suggest the most likely survivors are those clusters that are
either more highly concentrated (King model concentrations of greater
than 1.0-1.5), and/or those with steep IMFs.  This is shown graphically
in Fig.~9 where we have reproduced approximately a figure from
\cite{TPZ00} (their Figure 8) that shows which range of cluster
parameters the clusters are long-lived or disrupt.  For the range of
plausible IMF slopes that we have determined (assuming, as in the
models, that the IMF extends down to sub-solar masses), the models
suggest that clusters that are about as concentrated as we have observed
in the Antennae will survive for at least a few Gyrs.  Even though we
have no independent constraint on the lower mass limit of the stellar
mass function, it is interesting that models with steep IMF slopes are
also robust against disruption.  From these models alone it is difficult
to tell if the clusters are able to withstand the disruptive tendencies
of the large mass loss they suffer if the lower mass limit (0.1
M$_{\sun}$ in Figs.~6 and 7) is raised and the IMF slope steepened, such
that the observed light-to-mass ratio constraints are met.  The limited
number of models so far generated seems to indicate that clusters with
very steep slopes ($\alpha$ $>$ 3) of their IMF are robust.  More
modeling is obviously needed to answer this question quantitatively and
definitively.

\begin{figure}
\begin{minipage}{8.8cm}
\psfig{figure=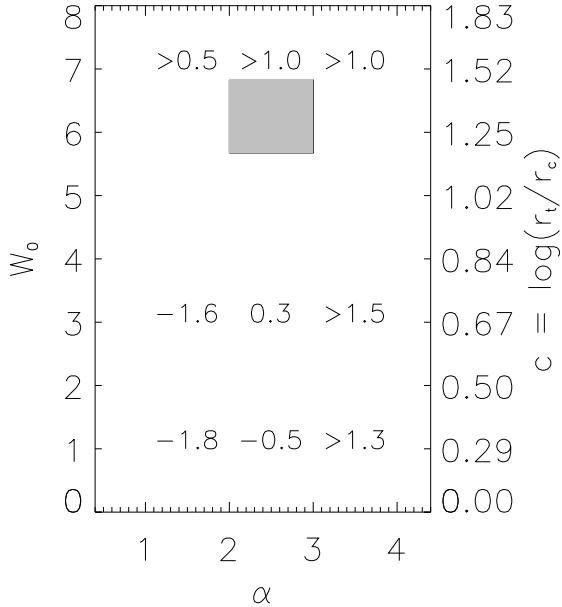,width=8.8cm}
\hspace{0cm}
\caption[Diagram of Cluster Survival]{This is a modified reproduction of
a figure from Takahashi \& Portegies Zwart (2000; their Figure 8)
showing the ages at the endpoint of evolution for modeled clusters.
The clusters were characterized by scaled central potentials, W$_0$
(which is related to the concentration, c, as shown in the ordinate on
the right side) and IMF slopes ($\alpha$).  The numbers in the grid give
the ages at the endpoint of the evolution (disruption or core collapse)
as log(age [Grys]).  Clusters with values below 0.3 (corresponding to 2
Gyrs) were disrupted by dynamical processes over time, the others
experienced core collapse.  The grey box shows the region where the
clusters in the Antennae are observed to lie.  This suggests that the
clusters are relatively long lived.}
\end{minipage}
\label{Clustersurvival}
\end{figure}

Even if the clusters survive, it is obviously important to have some
indication of how much mass loss they may undergo as they age.  For some
of the cluster parameters, these models suggest that these clusters may
lose significant amounts of mass ($>$50\%, sometimes much greater;
\cite{TPZ00}).  Clusters with steep IMF slopes ($\alpha$ $>$ 2.5) which
are compact (c$>$1.5) apparently suffer large mass loss (may lose more
than 90\% of their total initial mass; \cite{TPZ00} but also \cite{CW90}
whose models suggest that perhaps some of these high mass loss clusters
are disrupted).  Therefore, it is easily possible for these cluster to
survive, but over their lifetimes, they will have much lower masses than
when they are as young as the clusters we have observed here.  It is
through such strong mass loss that the young compact clusters in the
Antennae may evolve into something similar to the population of clusters
observed in galaxies generally (e.g., Fritze-v. Alvensleben 1998; 1999).

There are limitations in the models that might in fact lead to higher
mass loss rates and possibly easier disruption of clusters for a given
set of initial conditions than predicted in the models.  One is that the
tidal field is much more complex and stronger than what is assumed in
the models which is static or a simple orbital potential in a disk.  In
a strong and variable tidal field, the mass loss rates would undoubtedly
go up (e.g., Gnedin \& Ostriker 1997).  However, very little is known
about the tidal field over time on the scales of the clusters and such an
analysis would require a greater understanding of the mass distribution
in galaxies than we currently have.  Therefore, literal interpretations
of the evolution of clusters as indicated by the models should be viewed
with some caution.

In summary, given what we have learned in this analysis, it is likely
that the most massive clusters are able to survive in the galaxian
environment in which they are born, and do not abruptly disappear.
Since we cannot constrain the lower mass limit of the stellar mass
function it is possible that the clusters are easier to disrupt than we
have suggested here.  The amount of mass-loss they will undergo is
uncertain but is likely to be a large percentage of their initial masses
thereby providing a mechanism were the massive clusters studied here can
evolve to look more like a typical cluster in a galaxy like the Milky
Way.

\section{Summary and Conclusions}

In order to estimate the masses of young compact clusters in the
interacting pair of galaxies NGC 4038/4039 (``the Antennae"), we have
measured the velocity dispersions of 6 clusters using high resolution
optical and near-IR spectroscopy conducted at the ESO-VLT.  The velocity
dispersions were estimated using the stellar absorption features of CO
at 2.29 $\mu$m and metal absorption lines at around 8500\AA\ including
lines of the Calcium Triplet.  To estimate the masses, we measured
the size scales and light profiles from archival HST WFPC-2 (see
Whitmore et al.  1999).  The ages of individual clusters were estimated
using the equivalent widths of the CO band-head at 2.29 $\mu$m, the CaT,
and the Br$\gamma$ emission line (estimated from data in Mengel et al.
2001 and Mengel et al. 2001, in preparation).  Our principle results
from this analysis are:

\begin{itemize}

\item The clusters have ages of about 8 Myrs.  This is a selection
effect since we selected the clusters based on other observations
indicating that they had ages of about 10 Myrs to ensure they had strong
CO$\lambda$2.29 $\mu$m and CaT absorption lines for determining the
velocity dispersions.

\item The best fitting light profiles are consistent with King models of
concentration parameters, c, of either 1.176 or 1.477 (tidal to core
radius ratios of 15 or 30 respectively).  The half-light radii from
these fits are all around 4 pc which is typical of the young compact
clusters in the Antennae (Whitmore et al. 1999).

\item The measured velocity dispersions were all between 9 and 21 km
s$^{-1}$.  Combining the velocity dispersions, half-light radii, and
parameters of a King model under the assumption that the clusters are in
approximate Virial equilibrium (an assumption which is supported by our
estimate that the clusters have already survived for 10s of crossing times)
yields mass estimates that range from 6.5 $\times 10^5$ to 4.7 $\times
10^6$M$_{\sun}$.  These masses are large compared to typical masses of a
globular cluster ($\sim 1 \times 10^5$M$_{\sun}$), but this not
surprising, since we selected the brightest clusters for our first
analysis.

\item Comparing the cluster light-to-mass ratios with stellar synthesis
models (Leitherer et al. 1999) suggests that these clusters may exhibit
a range of initial mass functions, with clusters in the ``over-lap''
region showing evidence for a steeper IMF slope than those clusters in
less extinguished regions.  However, with only five clusters observed this
result needs to be substantiated with observations of more clusters.

\item Results of Fokker-Planck simulations of compact clusters with
concentrations and IMF parameters similar to those found for the young,
compact clusters in the Antennae studied here, suggest that these
clusters are likely to be long-lived (more than a few Gyrs) and may lose
a substantial fraction of their total mass.

\end{itemize}

\begin{acknowledgements}

We would like to thank Brad Whitmore for providing size estimates of the
clusters we observed, for interesting discussions concerning the
young clusters in the Antennae and for his insightful, to the point comments
when he refereed the paper; S\o ren Larsen for providing us with his
cluster fitting code; and Linda Tacconi for taking additional stellar
spectra used in this analysis.

\end{acknowledgements}

\end{document}